\documentstyle[12pt,epsfig]{article}
\def\gtrsim{\mathrel{\mathpalette\vereq>}}
\makeatletter
\def\vereq#1#2{\lower3pt\vbox{\baselineskip1.5pt \lineskip1.5pt
\ialign{$\m@th#1\hfill##\hfil$\crcr#2\crcr\sim\crcr}}}
\makeatother
\newcounter{axn}

\def\diag{{\rm diag}} 
\def\PR#1#2#3{Phys. Rev. {\bf #1} (#3) #2 }
\def\PRL#1#2#3{Phys. Rev. Lett. {\bf #1} (#3) #2 }
\def\PL#1#2#3{Phys. Lett. {\bf #1} (#3) #2 }

\def\NP#1#2#3{Nucl. Phys. {\bf #1} (#3) #2 }

\def\PTP#1#2#3{Prog. Theor. Phys. {\bf #1} (#3)#2 }

\begin{document}

\begin{titlepage}
\begin{flushright}
hep-ph/0401183\\
KUNS-1894 \\
%{DPNU-02-xx}\\
\today
\end{flushright}

\vskip 1cm
\begin{center}
{\large\bf Dynamical symmetry breaking in Gauge-Higgs unification \\ 
 on orbifold}
\vspace{0.5cm}

{\bf  Naoyuki Haba$^{1}$, Yutaka Hosotani$^{2}$, 
Yoshiharu Kawamura$^{3}$, \\ and Toshifumi Yamashita$^{4}$}\\ 
\vspace{0.5cm}
$^1${\small \it Institute of Theoretical Physics, 
 University of Tokushima, Tokushima 770-8502, Japan} \\
$^2${\small \it Department of Physics, Osaka University,
Toyonaka, Osaka 560-0043, Japan}\\
$^3${\small \it Department of Physics, Shinshu University,
Matsumoto, Nagano 390-8621, Japan}\\
$^4${\small \it Department of Physics, Kyoto University,
Kyoto, 606-8502, Japan}\\
\end{center}
\vskip .5cm
\vspace{.3cm}

%%%%%%%%%%%%%%%%%%%%%%%%%%%%%%%%%%%%%%%%%%%%%%
%%%%%%%%%%%%%%  ABSTRACT %%%%%%%%%%%%%%%%%%%%%
%%%%%%%%%%%%%%%%%%%%%%%%%%%%%%%%%%%%%%%%%%%%%%
\begin{abstract}

%\noindent
We study the dynamical symmetry breaking in the 
 gauge-Higgs unification of the 5D theory 
 compactified on an orbifold, $S^1/Z_2$. 
This theory identifies Wilson line degrees of freedoms 
 as ``Higgs doublets''. 
We consider $SU(3)_c \times SU(3)_W$ and 
 $SU(6)$ models with the compactification scale 
 of order a few TeV. 
The gauge symmetries are reduced 
 to $SU(3)_c \times SU(2)_L \times U(1)_Y$ and 
 $SU(3)_c \times SU(2)_L \times U(1)_Y \times U(1)$, 
 respectively, 
 through 
 the 
 orbifolding boundary conditions. 
We estimate the one loop effective potential 
 of ``Higgs doublets'', and find that the electro-weak breaking 
 is realized through the radiative corrections 
 when there are suitable numbers of bulk fields 
 possessing the suitable representations.  
The masses of ``Higgs doublets'' are 
 ${\mathcal O}(100)$ GeV in this scenario.

\end{abstract}
\end{titlepage}
\setcounter{footnote}{0}
\setcounter{page}{1}
\setcounter{section}{0}
\setcounter{subsection}{0}
\setcounter{subsubsection}{0}

%%%%%%%%%%%%%%%%%%%%%%%%%%%%%%%%%%%%%%%%%%%%%%%%%%%%%%%%%%%%%%%%%%%
%%%%%%%%%%%%%%%%%%%%% INTRODUCTION %%%%%%%%%%%%%%%%%%%%%%%%%%%%%%%%
%%%%%%%%%%%%%%%%%%%%%%%%%%%%%%%%%%%%%%%%%%%%%%%%%%%%%%%%%%%%%%%%%%%
\section{Introduction}

Much attentions have been paid to gauge 
 theories in higher dimensions\cite{5d,Kaplan,su3-2,HHHK,HHK}. 
(See, for examples, grand unified theories (GUTs) in 
 higher dimensions on 
 orbifolds.) 
One of the strongest motivations of 
 the higher dimensional gauge theory
 is based on the very 
 attractive idea that the gauge and the Higgs fields can be 
 unified in higher dimensions\cite{Manton:1979kb,YH}.
Recently, this possibility has been revisited in 
 Ref.\cite{Krasnikov:dt,Lim,Hall:2001zb,Burdman:2002se,HS,mimura1,Csaki,ph-gaugehiggs,26}.
In these scenarios the Higgs doublets 
 are identified with the extra-dimensional
 components of the gauge fields in higher dimensions.
The masses of ``Higgs fields'' are forbidden by 
 the higher dimensional gauge invariance. %, 
This is the reason why the 
 ``Higgs fields'' have at most only 
 finite masses of the order of the 
 compactification scale 
 in these scenarios. 
The gauge group in higher dimensions must be larger than the 
 standard model (SM) gauge
 group in order to obtain the 
 ``Higgs doublets'' from the gauge fields
 in higher dimensions. 
The gauge symmetries are reduced 
 by the orbifolding boundary conditions of extra 
 dimensions. 
The identification of ``Higgs fields'' as 
 a part of gauge super-multiplet has 
 been considered in 
 5D $N=1$ supersymmetric (SUSY)
 gauge theory 
 whose 5th coordinate is compactified
 on $S^1/Z_2$ orbifold\cite{Lim,Hall:2001zb,Burdman:2002se,HS,mimura1}, 
 which corresponds to the 4D $N=2$ SUSY gauge theory. 
Also it is 
 considered in 6D $N=2$ SUSY 
 gauge theory, 
 whose 5th and 6th coordinates are compactified
 on $T^2/(Z_2 \times Z_2')$ orbifold\cite{Hall:2001zb}, which  
 corresponds to the 4D $N=4$ SUSY gauge theory.

In this paper, we consider 
 5D non-SUSY and $N=1$ SUSY theories 
 compactified on an orbifold, $S^1/Z_2$, 
 where Wilson line degrees of freedoms 
 are identified as ``Higgs doublets''. 
Quarks and leptons are assumed to be localized 
 on the 4D wall. 
We consider $SU(3)_c \times SU(3)_W$ and 
 $SU(6)$ models with 
 the compactification scale being 
 a few TeV. 
The gauge symmetries are reduced 
 to $SU(3)_c \times SU(2)_L \times U(1)_Y$ and 
 $SU(3)_c \times SU(2)_L \times U(1)_Y \times U(1)$,
 respectively, 
 through 
 the 
 orbifolding boundary conditions. 
We estimate the one loop effective potential 
 of ``Higgs doublets'', and find that the electro-weak breaking 
 is realized through the radiative corrections 
 when there are suitable numbers of bulk fields 
 possessing the suitable representations.  
The masses of ``Higgs doublets'' are 
 ${\mathcal O}(100)$ GeV in this scenario.   
The suitable value of $\sin^2 \theta_W$ and the gauge coupling 
 unification are assumed to be realized 
 by the effects of wall-localized 
 kinetic terms, which may not respect 
 the bulk symmetry\footnote{
 For other possibilities, 
 the power law unification\cite{Dienes:1998vh}
 or the accelerated unification\cite{Arkani-Hamed:2001vr} 
 might be useful. 
}. 
We should also assume the baryon number symmetry 
 to avoid rapid proton decay in the TeV scale
 compactification.

%%%%%%%%%%%%%%%%%%%%%%%%%%%%%%%%%%%%%%%%%%%%%%%%%%%%%%%%%%%%%%%%%%%
%%%%%%%%%%%%%%%%%%%%% SECTION %%%%%%%%%%%%%%%%%%%%%%%%%%%%%%%%%%%%%
%%%%%%%%%%%%%%%%%%%%%%%%%%%%%%%%%%%%%%%%%%%%%%%%%%%%%%%%%%%%%%%%%%%
\section{Gauge-Higgs unification on $S^1/Z_2$}

We will consider 5D $SU(N)$ gauge 
 theory on $S^1/Z_2 \otimes M^{4}$. 
The gauge fields propagate in the bulk. 
The 5th dimensional coordinate $(y)$ 
 is assumed to be compactified on an $S^1/Z_2$ 
 orbifold. 
Under the parity transformation of $Z_2$ 
 which transforms $y \rightarrow -y$,  
 the gauge field $A_M(x^\mu,y)$ $(M= \mu( =0$-$3),5)$ 
 in the 5D space-time 
 transforms as 
\begin{eqnarray}
 && A_\mu(x^\mu,-y) = PA_\mu(x^\mu,y)P^{\dagger},\\
 && A_5(x^\mu,-y) = -PA_5(x^\mu,y)P^{\dagger},
\end{eqnarray}
where $P$ is the operator of 
 $Z_2$ transformation. 
Two walls at $y=0$ and $\pi R$ are
 fixed points under $Z_2$ transformation. 
The physical space can be taken to $0 \leq y \leq \pi R$. 
Considering the $S^1$ 
 boundary condition, 
$$
A_M(x^\mu,y+2\pi R) = TA_M(x^\mu,y)T^{\dagger},
$$
 the reflection 
 around $y=\pi R$, $Z_2'$,   
 is given by 
$$
P'=TP.
$$
The gauge field $A_M(x^\mu,y)$ 
 transforms 
\begin{eqnarray}
 && A_\mu(x^\mu,\pi R-y) = P'A_\mu(x^\mu, \pi R+y)P'^{\dagger},\\
 && A_5(x^\mu,\pi R-y)   = -P'A_5 (x^\mu, \pi R+y)P'^{\dagger}. 
\end{eqnarray}
under the parity transformation of $Z_2'$. 
It should be noticed that 
 the signs of parities of $A_5$ are opposite to
 those of $A_\mu$.  
According to eigenvalues, $(\pm, \pm)$, of 
 parities, $(P, P')$, 
 the field $A_\mu (x^\mu,y)$ is divided 
 into four eigenfunctions as 
\begin{eqnarray}
\label{33}
&& A_\mu (x^\mu,y)_{(+,+)}={1\over \sqrt{2^{\delta_{n,0}}\pi R}}
            \sum_{n=0}^{\infty}A_\mu^{(n)}(x^\mu)_{(+,+)}
                         \cos\left({ny\over R}\right), \\
&& A_\mu (x^\mu,y)_{(+,-)}={1\over \sqrt{\pi R}}
     \sum_{n=0}^{\infty}A_\mu^{(n)}(x^\mu)_{(+,-)}
     \cos\left({(n+1/2)y\over R}\right), \\
&& A_\mu (x^\mu,y)_{(-,+)}={1\over \sqrt{\pi R}}
     \sum_{n=0}^{\infty}A_\mu^{(n)}(x^\mu)_{(-,+)}
     \sin\left({(n+1/2)y\over R}\right), \\
&& A_\mu (x^\mu,y)_{(-,-)}={1\over \sqrt{\pi R}}
     \sum_{n=0}^{\infty}A_\mu^{(n)}(x^\mu)_{(-,-)}
     \sin\left({(n+1)y\over R}\right).
\label{66}
\end{eqnarray}
The expansion of $A_5 (x^\mu ,y)$ 
 is done in the same way. 
The parity eigenvalue of the field $A_5 (x^\mu,y)$, 
 is opposite to that of 
 $A_\mu$. 
The massless states surviving in the low energy 
 are zero mode components with parity transformation 
 $(P, P')=(+, +)$. 
This paper will consider the situation 
 that zero modes of $A_5$ are $({\bf1},{\bf2},1/2)$ or
 $({\bf1},{\bf2},-1/2)$ under $SU(3)_c \times SU(2)_L \times U(1)_Y$. 
We regard these components as ``Higgs doublets'', 
 then we call this theory gauge-Higgs unification. 
Here the local gauge invariance in the 5D guarantees 
 the masslessness of the ``Higgs field'', 
 so the Higgs mass should be finite 
 after the radiative corrections. 
We will study two models 
 in the following sections. 
In section 3, we consider 
 $SU(3)_c\times SU(3)_W$ 
 gauge theory, 
 where the nontrivial 
 parity operators, 
 $P=P'= \diag(1,1,-1)$ 
 realizes the gauge reduction of 
 $SU(3)_W \rightarrow SU(2)_L\times U(1)_Y$ 
 as well as the ``Higgs doublets'' 
 appear as the zero modes 
 in $A_5$\cite{Lim,Hall:2001zb,Burdman:2002se}. 
In section 4, 
 we will consider the 5D $SU(6)$ theory with
 the $Z_2$ parity operators, $P= \diag(1,1,1,1,-1,-1)$ and 
 $P'=\diag(1,-1,-1,-1,-1,-1)$\cite{Hall:2001zb,Burdman:2002se}. 
In both models, we estimate 
 the one loop effective potential 
 of the ``Higgs doublets'' including 
 the effects of 
 Kaluza-Klein (KK)\cite{KK} modes 
 and bulk matter fields.
And we study the vacuum structure 
 of the models and calculate 
 the mass of the ``Higgs fields''. 
We will also study the SUSY version, where
 $A_5$ becomes the imaginary part of
 an adjoint chiral superfield 
 after dimensional reduction 
 as follows. 
Since the 5D $N=1$ SUSY theory corresponds to 
 4D $N=2$ SUSY theory,  
 the 5D gauge multiplet, 
$$
{\mathcal{V}}=(A^M, \lambda, \lambda', \sigma),
$$
is decomposed to a vector super-field and an adjoint chiral 
 superfield as
$$
V=(A^\mu, \lambda), \;\;\;\;\;\;\;
\Sigma = (\sigma + iA^5, \lambda'),
$$ 
respectively. 
Then, in the SUSY case, 
 the gauge multiplet transforms as 
\begin{eqnarray}
&&\left(
\begin{array}{c}
 V (x^\mu, -y)\\
 \Sigma (x^\mu, -y)
\end{array}
\right) =
 P \left(
\begin{array}{c}
V (x^\mu, y)\\
 -\Sigma (x^\mu, y)
\end{array}
\right) P^{\dagger}, \\
&&\left(
\begin{array}{c}
 V (x^\mu, \pi R-y)\\
 \Sigma (x^\mu, \pi R-y)
\end{array}
\right) =
 P' \left(
\begin{array}{c}
V (x^\mu, \pi R+y)\\
 -\Sigma (x^\mu, \pi R+y)
\end{array}
\right) P'^{\dagger}, 
\label{P}
\end{eqnarray}
corresponding to Eqs.(1)-(4).

%%%%%%%%%%%%%%%%%%%%%%%%%%%%%%%%%%%%%%%%%%%%%%%%%%%%%%%%%%%%%%%%%%%
%%%%%%%%%%%%%%%%%%%%% SECTION %%%%%%%%%%%%%%%%%%%%%%%%%%%%%%%%%%%%%
%%%%%%%%%%%%%%%%%%%%%%%%%%%%%%%%%%%%%%%%%%%%%%%%%%%%%%%%%%%%%%%%%%%
\section{$SU(3)_c\times SU(3)_W$ model}

Let us study the possibility of the 
 dynamical symmetry breaking in the 
 $SU(3)_c \times SU(3)_W$ model, 
 where the Higgs doublets can be identified
 as the zero mode 
 components of 
 $A_5$\cite{Lim,Hall:2001zb,Burdman:2002se}. 
We take 
\begin{eqnarray}
&&P=P'=\diag(1,-1,-1) 
\end{eqnarray}
in the base of $SU(3)_W$\footnote{As for 
 $SU(3)_c$, we take $P=P'=I$.}. 
Then, they divide $A_\mu$ and $A_5$ as,
\begin{eqnarray}
&&A_\mu=\left(
\begin{array}{c|cc}
(+,+)&(-,-)&(-,-) \\ \cline{1-3}
(-,-)&(+,+)&(+,+) \\ 
(-,-)&(+,+)&(+,+) 
\end{array}
\right),\\
&&A_5 =\left(
\begin{array}{c|cc}
(-,-)&(+,+)&(+,+) \\ \cline{1-3}
(+,+)&(-,-)&(-,-) \\ 
(+,+)&(-,-)&(-,-) 
\end{array}
\right), 
\end{eqnarray}
which suggest 
 $SU(3)_W$ is broken down to 
 $SU(2)_L\times U(1)_Y$, 
 and there appear 
 one ``Higgs doublet'' in $A_5$ 
 as the zero mode\footnote{In SUSY case,
 there appear two ``Higgs doublets'' as 
 the zero modes.}. 
We assume that the compactification scale, 
 $R^{-1}$, as a few TeV.

The VEV of $A_5$ is written as 
\begin{equation}
\langle A_5 \rangle = 
   {1\over g R}\sum_a a_a {\lambda_a \over 2}, 
\label{A5}
\end{equation}
and we can always take VEV as 
$$
a_{1} =a,
$$
and $a_i=0$ for $i\neq1$
 by using the residual $SU(2)\times U(1)$ 
 {\it global} symmetry. 
The effective potential of $A_5$ 
 is given by\cite{Lim}
\begin{eqnarray}
V_{eff}^{gauge}
&=& -{3\over2}C\sum_{n=1}^{\infty}{1\over n^5}
    [\cos (2\pi na)+2\cos(\pi na)],
\label{Vg3}
\end{eqnarray}
where $C \equiv 3/(64\pi^7R^5)$. 
This means that the point at $a=0$ is the minimum 
 in $V_{eff}^{gauge}$, which suggests $SU(2)_L \times U(1)_Y$ is 
 not broken.

Then, in order to realize the 
 electro-weak symmetry breaking, let us 
 introduce extra fields in the bulk, 
which are 
 $N_s$ numbers of complex scalars, $\phi$, and 
 $N_f$ $(N_a)$ numbers of Dirac fermions, $\psi$ $(\psi^a)$, of the 
 fundamental (adjoint) 
 representation. 
They transform 
\begin{eqnarray}
\label{eta-s}
&&\phi(x, -y) = \eta P \phi(x, y) ~~, ~~ 
\phi(x, \pi R-y) = \eta' P' \phi(x, \pi R + y) , \\
\label{eta-D}
&&\psi(x, -y) = \eta P \gamma^5 \psi(x, y) ~~, ~~
\psi(x, \pi R-y) = \eta' P' \gamma^5 \psi(x, \pi R + y) ~~,\\
\label{eta-Da}
&&\psi^a(x, -y) = \eta P \gamma^5 \psi^a(x, y) P^{\dagger}~~, ~~
\psi^a(x, \pi R-y) = \eta' P' \gamma^5 \psi^a(x, \pi R + y) P'^{\dagger}~~,
\end{eqnarray}
under parities, 
 respectively. 
Here $\eta, \eta'=\pm$, and the effective potential 
 induced from these bulk fields strongly depends on 
 the sign of the product, $\eta \eta'$. 
Appendix B shows 
 the bulk fields' 
 contributions to the effective potential, 
\begin{eqnarray}
\label{Vm3}
V_{eff}^m
&=& C\sum_{n=1}^{\infty}{1\over n^5}
    [2 N_a^{(+)} \cos (2\pi na) + 
     2 N_a^{(-)} \cos (2\pi n(a-{1\over2})) \nonumber \\
&+& (4 N_a^{(+)}-N_s^{(+)} + 2 N_f^{(+)}) \cos(\pi na) \nonumber \\
&+& (4 N_a^{(-)}-N_s^{(-)} + 2 N_f^{(-)}) \cos(\pi n(a-1))]. 
\end{eqnarray}
The index $(\pm)$ indicates the
% the numbers of bulk fields with the 
 sign of $\eta \eta'$ in Eqs.(\ref{eta-s})-(\ref{eta-Da}). 
Here we denote 
 $N_s=N_s^{(+)}+N_s^{(-)}$, 
 $N_f=N_f^{(+)}+N_f^{(-)}$, and
 $N_a=N_a^{(+)}+N_a^{(-)}$. 
Seeing 
 the 1st derivative of 
 $V_{eff}=V_{eff}^{gauge}+V_{eff}^m$, 
 each term of ${\partial V_{eff}/ \partial a}$ 
 has a factor 
    $\sin(\pi na)$,
 which 
 means that the stationary points exist
 at least at $a=0$ and $a=1$\footnote{ 
The potential has the symmetry
 $V_{eff}(-a)=V_{eff}(a)$, 
 so that we should only check the region of $0\leq a \leq 1$.}. 
The difference of the heights 
 between two points is given by 
\begin{eqnarray}
&&V_{eff}(a=0)-V_{eff}(a=1)= \nonumber \\
&& \;\;\;2 [4(N_a^{(+)}-N_a^{(-)})+2(N_f^{(+)}-N_f^{(-)})-
     (N_s^{(+)}-N_s^{(-)})-3]C
 \sum_{n=1}^{\infty}
 {1 \over (2n-1)^5}.
\label{d2}
\end{eqnarray}
This means that the symmetric point of $a=0$ 
 becomes deeper as 
 the number of 
 bulk scalars with $\eta \eta'=+$ and 
 bulk fermions with $\eta \eta'=-$ increase.  
On the other hand, 
 the effects of 
 scalars with $\eta \eta'=-$ and 
 fermions with $\eta \eta'=+$ 
 make the height of 
 $a=1$ decrease.  
When $a=1$ point becomes the vacuum, 
 the Wilson loop 
 becomes 
\begin{eqnarray}
W_C&=&
\exp (ig \int_{0}^{2 \pi R}dy{1\over g R} a 
 {\lambda_{1}\over 2} ) \nonumber \\
&=&
\exp (ig {1\over g R}{\lambda_{1}\over 2}2\pi R)
=
\left(
\begin{array}{ccc}
-1 & & \\
   &-1& \\
   & &1 
\end{array}
\right),  
\label{WC1}
\end{eqnarray}
which suggests $SU(2)_L\times U(1)_Y$ is broken down to 
 $U(1)_{em}\times U(1)$. 
Since the VEV is ${\mathcal O}(R^{-1})$, which is a few TeV, 
 this case can not reproduce the correct 
 weak scale VEV.

\begin{figure}
\begin{center}
\epsfig{file=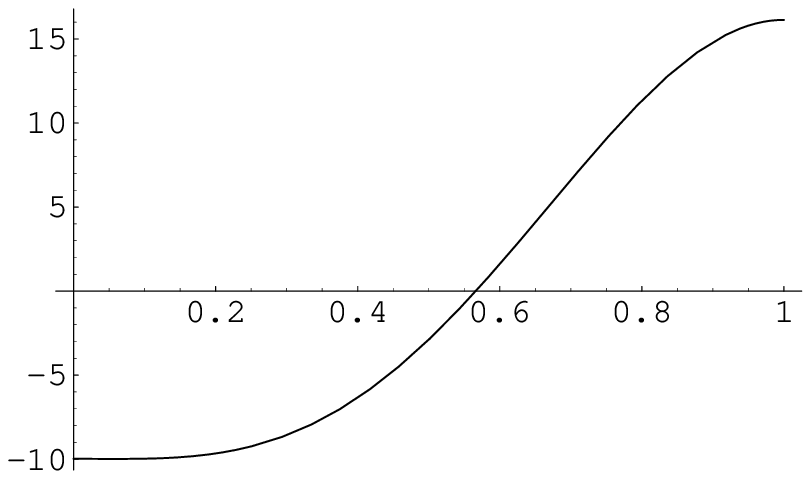,width=5cm}
\hspace{1cm}
\epsfig{file=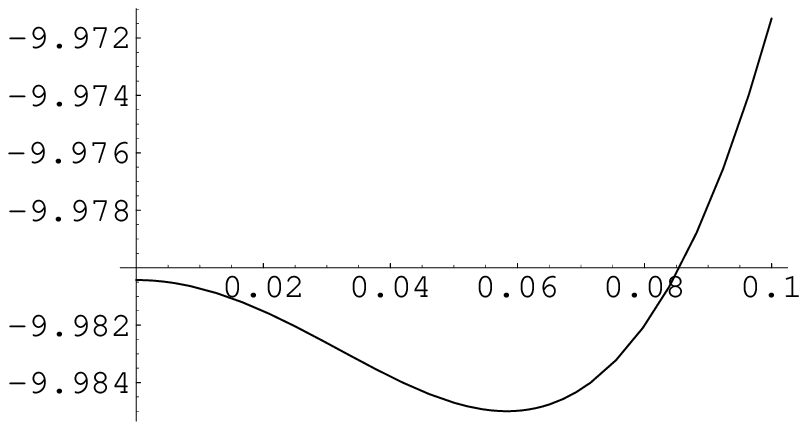,width=5cm}
\caption{The effective potential in the case of 
$N_a^{(+)}=2$, $N_f^{(-)}=8$, $N_s^{(+)}=4$, 
$N_s^{(-)}=2$, and $N_a^{(-)}=N_f^{(+)}=0$. 
The unit is $C=3/64\pi^7R^5$. 
The horizontal line shows $0\leq a \leq 1$ and $0\leq a \leq 0.1$. }
\label{fig:SU3}
\end{center}
\end{figure}

Anyhow, in order to realize the 
 suitable electro-weak symmetry breaking, 
 we must find another vacuum 
 at $(0<) a \ll 1$. 
In this case 
 $SU(2)_L\times U(1)_Y$ is broken down 
 to $U(1)_{em}$. 
Seeing $V_{eff}=V_{eff}^{gauge}+V_{eff}^m$ 
 in Eqs.(\ref{Vg3}) and (\ref{Vm3}), 
 we notice that 
 $n=1$ (of the summation of $n$) 
 has dominant contributions 
 for the form of the effective potential. 
Thus, we can obtain the suitable value 
 of $a$ $(a \ll 1)$,
 by introducing bulk fields which induce 
 large coefficients of $-\cos(\pi na)$ 
 and/or $\cos(\pi n(a-1))$, 
 and 
 small (but non-zero) coefficients of 
 $\cos(2\pi na)$ and $-\cos(2\pi n(a-1/2))$.  
We show an example which satisfies 
 the above condition, that is, 
 $N_a^{(+)}=2$, $N_f^{(-)}=8$, $N_s^{(+)}=4$, 
 $N_s^{(-)}=2$, and $N_a^{(-)}=N_f^{(+)}=0$.
Figure \ref{fig:SU3} %(\ref{fig:SU3a}) 
 shows the 
 $V_{eff}$ in the 
 region of $0\leq a \leq 1$ %($0\leq a \leq 0.1$). 
 and $0\leq a \leq 0.1$. 
The minimum exists at $a=0.058$, which is around
 the suitable magnitude of the weak scale in 
 TeV scale compactification. 
To be more precise, 
 the kinetic term of the ``Higgs field'' 
 is obtained from the 5D gauge kinetic term,
\[
 2\times\int dy\frac{1}{4}F_{\mu5}^aF^{a\mu5}
  =\int dy\frac{1}{2}\left( 
 \partial_\mu A_5^a + ig f^a_{bc}A_\mu^b A_5^c \right)^2
  =\left|\left( \partial_\mu 
   + ig_4 W^\alpha_\mu {\tau^\alpha \over 2} + 
    i\sqrt3g_4 {B_\mu\over2} 
         \right)H\right|^2.
\]
Where $H={\sqrt{2\pi R}}((A_5^1+iA_5^2)/\sqrt2, 
 (A_5^4+iA_5^5)/\sqrt2)^T$ is the 4D ``Higgs doublet'', 
 and $g_4$ is the 
 4D gauge coupling constant defined 
 as $g_4 = g/\sqrt{2\pi R}$.
This yields too large $\sin\theta_W$ and it seems hard 
 to reconcile it with the experimental value 
 through the renormalization group effect. 
However, as discussing in Ref.\cite{Burdman:2002se}, 
 4D gauge couplings can be also 
 affected by wall-localized 
 gauge kinetic terms, such as 
 $\delta(0)\lambda_0 F^{\mu\nu}{}^2$, 
 which do not respect the bulk symmetry.
When these couplings dominate 
 bulk gauge couplings\footnote{
For this situation, we need two assumptions. 
One is that the 
 wall-localized Higgs kinetic 
 term is negligibly small comparing 
 to the bulk Higgs kinetic term. 
The other is 
 the 
 bulk induced gauge coupling should be 
 larger than wall-localized ``gauge
 coupling'' as, 
 $g_4^2 > \lambda_0^{-1}$.
Thus, we should take $g_4 \gtrsim 1$, since 
 the wall-localized ``gauge couplings'', 
 such as $\lambda_0$, can mainly reproduce 
 the magnitudes of low energy 
 gauge couplings 
 of $SU(2)_L \times U(1)_Y$, $(g_2, g_Y)$, 
 under this situation.
And we simply assume $g_4={\mathcal O}(1)$ 
 in the following discussions.},  
 we can expect to have the suitable gauge couplings
 in the low energy.
In this case, 
 the normalization of the low 
 energy gauge fields might be changed as 
 $(W_\mu,B_\mu)\rightarrow(g_2/g_4W_\mu,
 g_Y/(\sqrt3g_4)B_\mu)$, which yields
 the usual Higgs kinetic term. 
Thus, we set
\[
  {\sqrt{2\pi R}}\left<A_5^1\right>
 =\frac{a}{g_4R}\sim246{\rm GeV}.
\]
%
%}%
The mass squared of the ``Higgs field'' is given by 
\begin{equation} 
m_{A_5}^2 = (gR)^2 
     \left.{\partial^2 V_{eff} \over \partial a^2}\right|_{a=0.058}
 = {3g_4^2 \over 32 \pi^4 R^2}
     \left.{\partial^2 (V_{eff}/(C\pi^2)) \over \partial a^2}\right|_{a=0.058}.
\end{equation}
By using the approximation formula,
\begin{eqnarray}
&&\sum^{\infty}_{n=1} 
{\cos(n\pi \xi) \over n^3}\simeq 
 \zeta(3)+{(\pi \xi)^2 \over 2}ln(\pi \xi)-{3\over4}(\pi \xi)^2, \\
&&\sum^{\infty}_{n=1} 
{\cos(n\pi (\xi-1)) \over n^3}\simeq 
 -{3\over4}\zeta(3)+{(\pi \xi)^2 \over 2}ln2, 
\end{eqnarray}
for a small $\xi$, 
 the ``Higgs'' mass becomes 
\begin{equation} 
m_{A_5}^2 \sim \left({0.031 g_4 \over R}\right)^2 
   \sim (130 g_4^2 \; {\rm GeV})^2, 
\end{equation}
where $g_4 = {\mathcal O}(1)$.

Now let us consider the SUSY case. 
Since the effective potential is zero when 
 SUSY remains,  
 we adopt Scherk-Schwarz (SS) SUSY 
 breaking\cite{SS,SS2,SS3,SS4}\footnote{For
 the effective potential in 
 other SUSY breaking, the calculation 
 has benn done in the case of the 
 $S^1$ compactification in 
 Ref.\cite{Takenaga:2003dp}.
 }, 
 where the mode expansions are 
 given in Ref.\cite{HHHK}. 
The effective potential 
 in the SUSY version is given by 
\begin{eqnarray}
V_{eff}^{gauge}
&=& -2 C \sum_{n=1}^{\infty}{1\over n^5}
    (1-\cos(2\pi n\beta)) [\cos(2 \pi n a)+2\cos(\pi n a)],
\label{26}
\end{eqnarray}
where $\beta$ parameterizes SS SUSY breaking. 
We take $\beta \sim 0.1$, since 
 the soft mass is given by 
 $\beta /R$\cite{HHHK}. 
Since $(1-\cos(2\pi n\beta))\geq 1$, 
 Eq.(\ref{26}) means 
 that $a=0$ is the minimum point 
 in the effective potential.
Therefore, 
 the $SU(2)_L \times U(1)_Y$ is
 not broken
 as in the non-SUSY case.

What is going on 
 if the extra fields exist in the bulk? 
We take the bulk fields of 
 $N_f$ and $N_a$ species of 
 hypermultiplets of fundamental $(\Psi)$ and 
 adjoint $(\Psi^a)$ representations, respectively. 
The 
 bulk hypermultiplets, 
 ${\Psi}=(\phi, \phi^c{}^\dagger, \tilde{\phi}, 
 \tilde{\phi}^c{}^\dagger)$ 
 and 
${\Psi}^a=(\phi^a, \phi^a{}^c{}^\dagger, \tilde{\phi^a}, 
 \tilde{\phi^a}{}^c{}^\dagger)$, 
which 
 are decomposed into chiral superfields as, 
 $\Phi=(\phi, \tilde{\phi})$, 
 $\Phi^a=(\phi^a, \tilde{\phi^a})$, 
 and 
 $\Phi^c=(\phi^c, \tilde{\phi}^c)$, 
 $\Phi^a{}^c=(\phi^a{}^c, \tilde{\phi^a}{}^c)$, 
 where $\Phi$ (${\Phi^a}$) and 
 $\Phi^c$ (${\Phi^a{}^c}$) 
 have conjugated transformation 
 under the gauge group. 
They transform 
\begin{eqnarray}
\label{22}
&&\left(
\begin{array}{c}
 \Phi (x^\mu, -y) \\ 
 \Phi^c{}^\dagger (x^\mu, -y) 
\end{array}
\right) =
 \eta P \left(
\begin{array}{cc}
\Phi (x^\mu, y) \\ 
 -\Phi^c{}^\dagger (x^\mu, y) 
\end{array}
\right) , \\
&&\left(
\begin{array}{c}
 \Phi (x^\mu, \pi R-y) \\ 
 \Phi^c{}^\dagger (x^\mu, \pi R-y) 
\end{array}
\right) =
 \eta' P' \left(
\begin{array}{cc}
\Phi (x^\mu, \pi R+y) \\ 
 -\Phi^c{}^\dagger (x^\mu, \pi R+y) 
\end{array}
\right) ,
\label{23} \\
\label{24}
&&\left(
\begin{array}{c}
 \Phi^a (x^\mu, -y) \\ 
 \Phi^a{}^c{}^\dagger (x^\mu, -y) 
\end{array}
\right) =
 \eta P \left(
\begin{array}{cc}
\Phi^a (x^\mu, y) \\ 
 -\Phi^a{}^c{}^\dagger (x^\mu, y) 
\end{array}
\right) P^\dagger, \\
&&\left(
\begin{array}{c}
 \Phi^a (x^\mu, \pi R-y) \\ 
 \Phi^a{}^c{}^\dagger (x^\mu, \pi R-y) 
\end{array}
\right) =
 \eta' P' \left(
\begin{array}{cc}
\Phi^a (x^\mu, \pi R+y) \\ 
 -\Phi^a{}^c{}^\dagger (x^\mu, \pi R+y) 
\end{array}
\right) P'^\dagger,
\label{25}
\end{eqnarray}
under the parities, 
 respectively. 
According to the sign of $\eta \eta'$ 
 we denote 
 $N_f=N_f^{(+)}+N_f^{(-)}$ and 
 $N_a=N_a^{(+)}+N_a^{(-)}$. 
We always take even number of $N_f^{(\pm)}$ 
 to avoid the gauge anomaly. 
Appendix B suggests 
 the extra matter contributions for the effective potential 
 are given by
\begin{eqnarray}
V_{eff}^m
&=& 2 C\sum_{n=1}^{\infty}{1\over n^5}
         (1-\cos(2\pi n\beta)) \times 
 [N_a^{(+)}\cos(2\pi na)+N_a^{(-)}
  \cos(2\pi n(a-{1\over2}))\nonumber \\
&& +   2(N_a^{(+)}+{N_f^{(+)} \over 2}) \cos(\pi na)+
       2(N_a^{(-)}+{N_f^{(-)} \over 2}) \cos(\pi n(a-1))]. 
\label{Vm}
\end{eqnarray}
As the non-SUSY case, 
 the 1st derivative of $V_{eff}=V_{eff}^{gauge}+V_{eff}^m$ 
 has the factor $\sin(\pi na)$, 
 which means 
 the stationary points exist
 at $a=0$ and $a=1$. 
The difference of the heights 
 between two points is given by 
\begin{eqnarray}
&&V_{eff}(a=0)-V_{eff}(a=1)= \nonumber \\
&& 
-8[1-(N_a^{(+)}-N_a^{(-)})
 -{(N_f^{(+)}-N_f^{(-)}) \over 2}] C 
  \sum_{n=1}^{\infty}
 {1 \over (2n-1)^5}
 (1-\cos(2\pi (2n-1)\beta)) .
\label{d1}
\end{eqnarray}
This means that 
 the height of 
 the point at $a=0$ becomes high (low) as 
 increasing the number of $N_a^{(+)}$ ($N_a^{(-)}$)
 and $N_f^{(+)}$ ($N_f^{(-)}$). 
Equation (\ref{d1}) is consistent with 
 the results in Ref.\cite{HHK}. 
As for the case of $N_a^{(+)}=1$,
 $N_f^{(+)}=N_f^{(-)}=N_a^{(-)}=0$, 
 the effective potential vanishes as $V_{eff}\equiv 0$, 
 since this case has 5D $N=2$ SUSY and 
 there is the residual SUSY after the 
 SS SUSY breaking\cite{Takenaga:2001re}.

\begin{figure}
\begin{center}
\epsfig{file=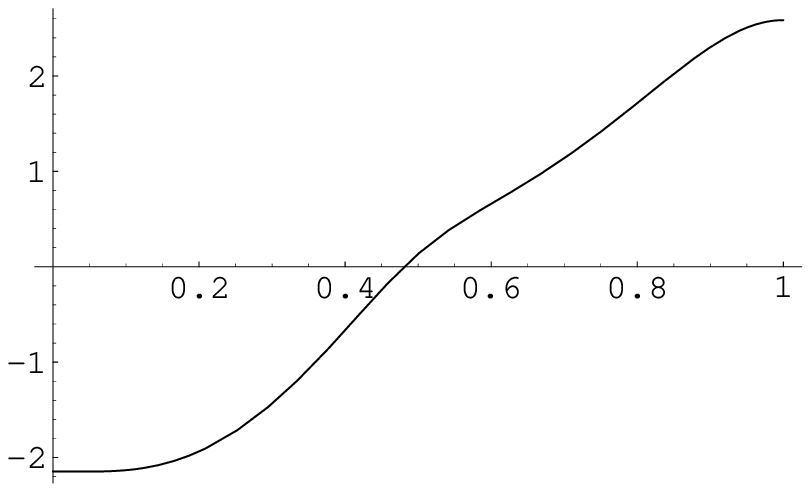,width=5cm}
\hspace{1cm}
\epsfig{file=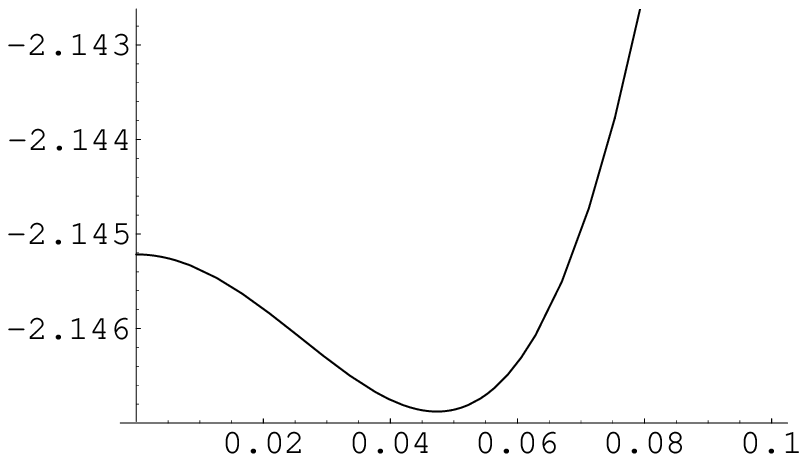,width=5cm}
\caption{The effective potential in the case of 
$N_a^{(+)}=N_a^{(-)}=2$, $N_f^{(-)}=4$, $N_f^{(+)}=0$
 with $\beta=0.1$.
The unit is $C=3/64\pi^7R^5$. 
The horizontal line shows $0\leq a \leq 1$ and $0\leq a \leq 0.1$. }
\label{fig:SUSYSU3}
\end{center}
\end{figure}

As the non-SUSY case, 
 in order to obtain the suitable value 
 of $a$ $(a \ll 1)$,
 we should introduce bulk hypermultiplets which induce 
 large coefficients of $-\cos(\pi na)$ and/or $\cos(\pi n(a-1))$ 
 and 
 small (but non-zero) coefficients of 
 $\cos(2\pi na)$ and $-\cos(2\pi n(a-1/2))$.  
We show an example which satisfies 
 the above condition, that is, 
 $N_a^{(+)}=N_a^{(-)}=2$, $N_f^{(-)}=4$, $N_f^{(+)}=0$
 (Fig.\ref{fig:SUSYSU3}). 
In this case with $\beta=0.1$, the minimum exists at 
 $a=0.047$, and the ``Higgs'' mass squared
 is given by 
\begin{equation} 
m_{A_5}^2 \sim \left({0.025 g_4 \over R}\right)^2
 \sim (130 g_4^2 \; {\rm GeV})^2,
\end{equation}
where $g_4 = {\mathcal O}(1)$.
It should be noticed that 
 the numerical analysis 
 shows 
 the value of $a$ depends on 
 that of $\beta$.

%%%%%%%%%%%%%%%%%%%%%%%%%%%%%%%%%%%%%%%%%%%%%%%%%%%%%%%%%%%%%%%%%%%
%%%%%%%%%%%%%%%%%%%%% SECTION %%%%%%%%%%%%%%%%%%%%%%%%%%%%%%%%%%%%%
%%%%%%%%%%%%%%%%%%%%%%%%%%%%%%%%%%%%%%%%%%%%%%%%%%%%%%%%%%%%%%%%%%%
\section{$SU(6)$ model}

We study the 
 vacuum structure 
 of the $SU(6)$ GUT, 
 in which 
 the Higgs doublets can be identified
 as the zero mode 
 components of $A_5$\cite{Hall:2001zb,Burdman:2002se}. 
We take 
\begin{eqnarray}
&&P=\diag(1,1,1,1,-1,-1), \\
&&P'=\diag(1,-1,-1,-1,-1,-1),
\end{eqnarray}
which divide $A_\mu$ and $A_5$ as
\begin{eqnarray}
&&A_\mu=\left(
\begin{array}{c|ccc|cc}
(+,+)&(+,-)&(+,-)&(+,-)&(-,-)&(-,-)\\ \cline{1-6}
(+,-)&(+,+)&(+,+)&(+,+)&(-,+)&(-,+)\\ 
(+,-)&(+,+)&(+,+)&(+,+)&(-,+)&(-,+)\\
(+,-)&(+,+)&(+,+)&(+,+)&(-,+)&(-,+)\\ \cline{1-6}
(-,-)&(-,+)&(-,+)&(-,+)&(+,+)&(+,+)\\
(-,-)&(-,+)&(-,+)&(-,+)&(+,+)&(+,+)
\end{array}
\right),\\
&&A_5 =\left(
\begin{array}{c|ccc|cc}
(-,-)&(-,+)&(-,+)&(-,+)&(+,+)&(+,+)\\ \cline{1-6}
(-,+)&(-,-)&(-,-)&(-,-)&(+,-)&(+,-)\\
(-,+)&(-,-)&(-,-)&(-,-)&(+,-)&(+,-)\\
(-,+)&(-,-)&(-,-)&(-,-)&(+,-)&(+,-)\\ \cline{1-6}
(+,+)&(+,-)&(+,-)&(+,-)&(-,-)&(-,-)\\
(+,+)&(+,-)&(+,-)&(+,-)&(-,-)&(-,-)
\end{array}
\right).
\end{eqnarray}
They suggest that 
 $P$ and $P'$ make $SU(6)$ broken to 
 $SU(3)_c\times SU(2)_L\times U(1)_Y \times U(1)$. 
Also, there appear 
 one ``Higgs doublet'' in $A_5$  
 as the zero mode. 
As in the previous section, 
 the compactification
 scale is assumed to be of order a few TeV. 
The VEV of $A_5$ is written in the similar
 way as Eq.(\ref{A5}) as 
\begin{equation}
\langle A_5 \rangle = 
   {1\over g R}a_{} {\lambda_{16} \over 2}
\end{equation}
by using the residual $SU(2)\times U(1)$ {\it global} symmetry. 
The calculation in the appendix A suggests 
 the gauge part of the 
 effective potential in the $SU(6)$ model 
 as
\begin{eqnarray}
V_{eff}^{gauge}
&=& -{3\over2}C\sum_{n=1}^{\infty}{1\over n^5}
    [6 \cos (\pi n (a-1))+ 2 \cos(\pi na)+\cos(2\pi na)]. 
\end{eqnarray}
{}From this equation, we can easily 
 show 
 that 
\begin{eqnarray}
&&V_{eff}^{gauge}(a=0)-V_{eff}^{gauge}(a=1)=12 \: C
 \sum_{n=1}^{\infty}
 {1 \over (2n-1)^5} \; >0. 
\end{eqnarray}
The numerical calculation of the effective 
 potential shows us 
 that $a=1$ ($a=0$) point is the global (local) minimum. 
The vacuum at $a=1$ has the 
 Wilson loop 
\begin{eqnarray}
W_C&=&
\exp (ig \int_{0}^{2 \pi R}dy{1\over g R} a 
 {\lambda_{16}\over 2} ) \nonumber \\
&=&
\exp (ig {1\over g R}{\lambda_{16}\over 2}2\pi R)
=
\left(
\begin{array}{cccccc}
-1 & & & & & \\
   &1& & & & \\
   & &1& & & \\
   & & &1& & \\
   & & & &-1 & \\
   & & & & &1 
\end{array}
\right),  
\label{WC2}
\end{eqnarray}
which suggests that $SU(2)_L\times U(1)_Y$ is broken down to 
 $U(1)_{em}\times U(1)$. 
This means that 
 the vacuum at $a=1$ is not suitable, 
 since the weak scale becomes too large. 
It can be possible to assume that 
 we exist at $a=0$ in the early universe, 
 and 
 the life time of this false vacuum 
 is longer than the universe history\cite{HHHK}. 
However, the suitable electro-weak 
 symmetry breaking can not be realized, 
 anyway.

Then, let us introduce 
 the extra fields in the bulk 
 for the suitable dynamical symmetry
 breaking of $SU(2)_L \times U(1)_Y$. 
We introduce the bulk fields of 
 $N_s$ numbers of complex scalars, $\phi$, and 
 $N_f$ $(N_a)$ numbers of Dirac fermions, $\psi$ $(\psi^a)$, of the 
 fundamental (adjoint) 
 representation. 
The transformations under parities 
 are the same as Eqs.(\ref{eta-s})-(\ref{eta-Da}). 
Appendix B suggests 
 that the effective potential induced from 
 the bulk fields is given by 
\begin{eqnarray}
V_{eff}^{m}
&=& C\sum_{n=1}^{\infty}{1\over n^5}[
    2N_a^{(+)}\cos(2\pi na)+
    2N_a^{(-)}\cos(2\pi n(a-{1\over2})) \nonumber \\ 
&& +(4N_a^{(+)}+12N_a^{(-)}+2N_f^{(+)}-N_s^{(+)})
     \cos(\pi na) \nonumber \\ 
&& +(12N_a^{(+)}+4N_a^{(-)}+2N_f^{(-)}-N_s^{(-)})\cos(\pi n(a-1))].
\end{eqnarray}
The index $(\pm)$ shows 
 the numbers of bulk fields with the 
 sign of $\eta \eta'$ in Eqs.(\ref{eta-s})-(\ref{eta-Da}). 
We denote as 
 $N_s=N_s^{(+)}+N_s^{(-)}$, 
 $N_f=N_f^{(+)}+N_f^{(-)}$, and
 $N_a=N_a^{(+)}+N_a^{(-)}$. 
As in the previous section, 
 the 1st derivative of $V_{eff}=V_{eff}^{gauge}+V_{eff}^m$ 
 suggests 
 that the stationary points exist
 at $a=0$ and $a=1$. 
The difference of the heights 
 between two points is given by 
\begin{eqnarray}
&&V^{eff}(a=0)-V^{eff}(a=1)= \nonumber \\
&&\;\;\;\;\;
2[6-8(N_a^{(+)}-N_a^{(-)})+
               2(N_f^{(+)}-N_f^{(-)})-(N_s^{(+)}-N_s^{(-)})]C
 \sum_{n=1}^{\infty}{1\over (2n-1)^5}. 
\label{d3}
\end{eqnarray}
The symmetric point of $a=0$ 
 becomes deeper (higher) as 
 $N_a^{(+)}$, $N_f^{(-)}$, $N_s^{(+)}$
 ($N_a^{(-)}$, $N_f^{(+)}$, $N_s^{(-)}$)
 increase.

\begin{figure}
\begin{center}
\epsfig{file=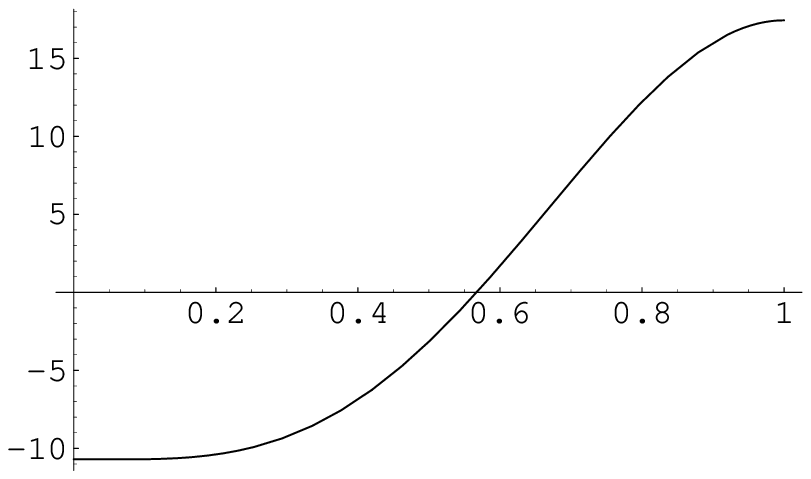,width=5cm}
\hspace{1cm}
\epsfig{file=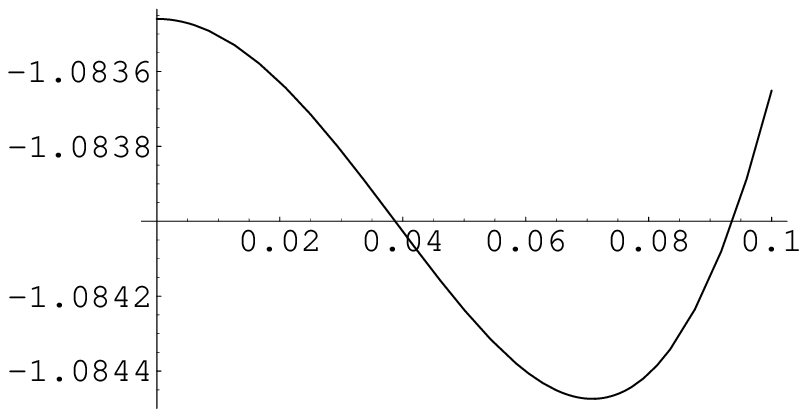,width=5cm}
\caption{The effective potential in the case of 
$N_a^{(+)}=N_f^{(-)}=2$, and others being zero. 
The unit is $C=3/64\pi^7R^5$. 
The horizontal line shows $0\leq a \leq 1$ and $0\leq a \leq 0.1$. }
\label{fig:SU6}
\end{center}
\end{figure}

As the previous section, 
 in order to obtain the suitable value 
 of $a$ $(a \ll 1)$,
 we should introduce bulk fields which induce 
 large coefficients of $-\cos(\pi na)$ and/or $\cos(\pi n(a-1))$ 
 and 
 small (but non-zero) coefficients of 
 $\cos(2\pi na)$ and $-\cos(2\pi n(a-1/2))$.  
We show an example which satisfies 
 the above condition, that is, 
 $N_a^{(+)}=N_f^{(-)}=2$, and others being 
 zero (Fig.\ref{fig:SU6}). 
In this case, the minimum exists at 
 $a=0.072$, and the ``Higgs'' mass squared
 is given by 
\begin{equation} 
m_{A_5}^2 \sim \left({0.038 g_4 \over R}\right)^2
 \sim (130 g_4^2 \; {\rm GeV})^2,
\end{equation}
where $g_4 = {\mathcal O}(1)$.

The effective potential 
 in the SUSY case is given by 
\begin{eqnarray}
V_{eff}^{gauge}
&=& -{2}C\sum_{n=1}^{\infty}{1\over n^5}
    (1-\cos(2\pi n\beta))\times \\
& & [6\cos (\pi n (a-1))+2\cos(\pi
    na)+\cos(2\pi na)]  \nonumber
\end{eqnarray}
{}from the calculation of the 
 appendix A. 
This shows 
\begin{equation}
V_{eff}^{gauge}(a=0)-V_{eff}^{gauge}(a=1)=16 \: C
 \sum_{n=1}^{\infty}
 {1 \over (2n-1)^5}(1-\cos(2\pi (2n-1)\beta)) \ >0, 
\end{equation}
and the numerical calculation really shows 
 $a=1$ is the global minimum. 
Then, in order to realize 
 the suitable electro-weak symmetry breaking, 
 we introduce 
 the extra hypermultiplets in the bulk 
 as in the previous section. 
We introduce $N_f$ ($N_a$) numbers of 
 fundamental (adjoint) 
 hypermultiplets, which 
 transform Eqs.(\ref{22}) and (\ref{23}) 
 (Eqs.(\ref{24}) and (\ref{25})) 
 under parities with 
 $\eta \eta'=\pm$. 
Appendix B shows that the bulk fields induce 
 the effective potential, 
\begin{eqnarray}
V_{eff}^{m}&=&
 {2}C\sum_{n=1}^{\infty}{1\over n^5}
    (1-\cos(2\pi n\beta))
   [N_a^{(+)}\cos (2\pi na) 
+    N_a^{(-)}\cos (2\pi n(a-{1\over2})) \nonumber \\
&&\;\;\;\;\;\;\;\;+
   (2N_a^{(+)}+6N_a^{(-)}+N_f^{(+)})\cos(\pi na) \nonumber \\
&&\;\;\;\;\;\;\;\;+
   (6N_a^{(+)}+2N_a^{(-)}+N_f^{(-)})\cos(\pi n(a-1))].
\end{eqnarray}
We can show 
 that the effective potential 
 vanishes in the case of $N_a^{(+)}=1$, 
 $N_f^{(+)}=N_f^{(-)}=N_a^{(-)}=0$ 
 due to the residual SUSY\cite{Takenaga:2001re}. 
The 1st derivative of $V_{eff}=V_{eff}^{gauge}+V_{eff}^m$ 
 suggests the stationary points exist
 at $a=0$ and $a=1$. 
The difference of the heights 
 between two points is given by 
\begin{eqnarray}
&&V_{eff}(0)-V_{eff}(1)= \nonumber\\
&& \hspace{-0.5cm} 2(8-8(N_a^{(+)}-N_a^{(-)})+2(N_f^{(+)}-N_f^{(-)})) 
   C\sum_{n=1}^{\infty}
 {1 \over (2n-1)^5}(1-\cos(2\pi (2n-1)\beta)),
\label{d5}
\end{eqnarray}
which is consistent with 
 the results in Ref.\cite{HHK}.

\begin{figure}
\begin{center}
\epsfig{file=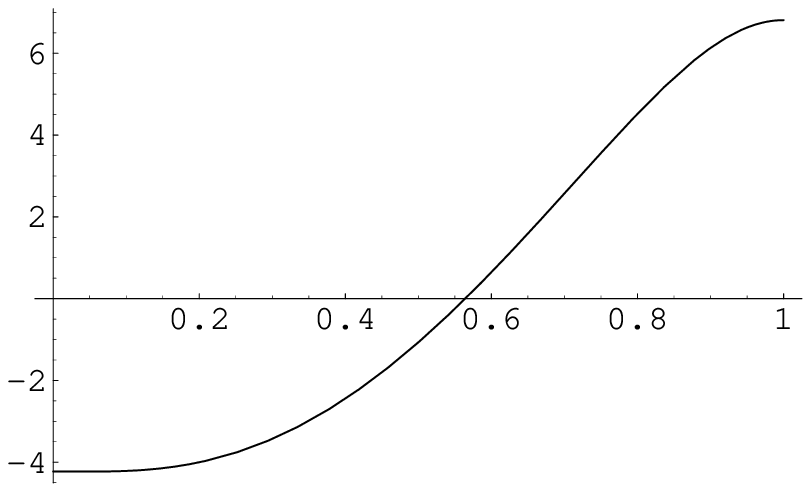,width=5cm}
\hspace{1cm}
\epsfig{file=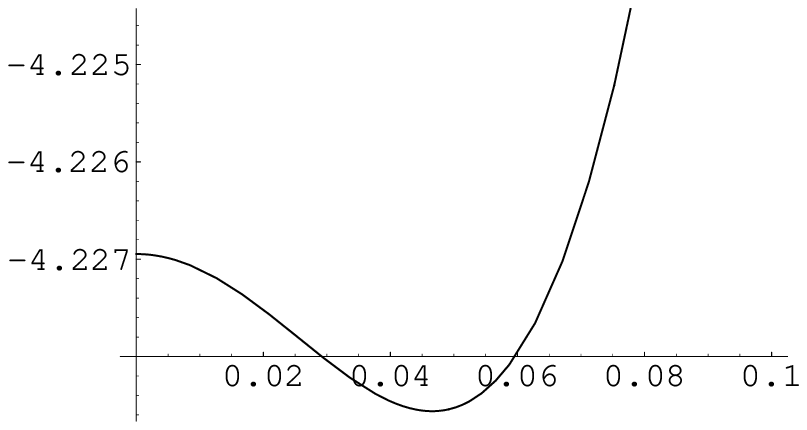,width=5cm}
\caption{The effective potential in the case of 
$N_a^{(+)}=2$, $N_f^{(-)}=10$, $N_a^{(-)}=N_f^{(+)}=0$
 with $\beta=0.1$.
The unit is $C=3/64\pi^7R^5$. 
The horizontal line shows $0\leq a \leq 1$ and $0\leq a \leq 0.1$. }
\label{fig:SUSYSU6}
\end{center}
\end{figure}

As the non-SUSY case, 
 in order to obtain the suitable value 
 of $a$ $(a \ll 1)$,
 we should introduce bulk hypermultiplets which induce 
 large coefficients of $-\cos(\pi na)$ and/or $\cos(\pi n(a-1))$ 
 and 
 small (but non-zero) coefficients of 
 $\cos(2\pi na)$ and $-\cos(2\pi n(a-1/2))$. 
We show an example which satisfies 
 the above condition, that is, 
 $N_a^{(+)}=2$, $N_f^{(-)}=10$, $N_a^{(-)}=N_f^{(+)}=0$
 with $\beta=0.1$ (Fig.\ref{fig:SUSYSU6}).  
In this case, the minimum exists at 
 $a=0.047$, and the ``Higgs'' mass squared
 is given by 
\begin{equation} 
m_{A_5}^2 \sim \left({0.024 g_4 \over R}\right)^2
 \sim (130 g_4^2 \; {\rm GeV})^2.
\end{equation}
where $g_4 = {\mathcal O}(1)$.

As for the extra $U(1)$ gauge symmetry that remains unbroken 
through the orbifolding boundary conditions,
 we assume it is broken by 
 an extra elementally Higgs field.

%%%%%%%%%%%%%%%%%%%%%%%%%%%%%%%%%%%%%%%%%%%%%%%%%%%%%%
\section{Summary and discussion}
%%%%%%%%%%%%%%%%%%%%%%%%%%%%%%%%%%%%%%%%%%%%%%%%%%%%%%

We have studied the possibility of the dynamical 
 symmetry breaking in the 
 gauge-Higgs unification in the 5D theory 
 compactified on an orbifold, $S^1/Z_2$. 
This theory identifies Wilson line degrees of freedoms 
 as ``Higgs doublets''. 
We considered $SU(3)_c \times SU(3)_W$ and 
 $SU(6)$ models with the compactification scale 
 of order a few TeV. 
The gauge symmetries are reduced 
 to $SU(3)_c \times SU(2)_L \times U(1)_Y$ and 
 $SU(3)_c \times SU(2)_L \times U(1)_Y \times U(1)$,
 respectively, 
 through 
 the 
 orbifolding boundary conditions. 
We have estimated the one loop effective potential 
 of ``Higgs doublets'', and find that the 
 electro-weak symmetry breaking 
 is realized through the radiative corrections 
 when there are suitable numbers of bulk fields 
 possessing the suitable representations.  
The masses of ``Higgs doublets'' are 
 ${\mathcal O}(100)$ GeV in this scenario.

Finally we should give a comment 
 on the Yukawa interactions. 
We have assumed that 
 quarks and leptons are localized 
 on the 4D wall. 
In this situation, 
 the ``Higgs doublet'' can not 
 make the gauge invariant Yukawa interactions 
 even if the ``Higgs doublet'' 
 appear as the zero mode of $\Sigma$
 as is shown in Ref.\cite{Hall:2001zb}. 
It is because 
 $\Sigma$ transforms 
 as $\Sigma \rightarrow e^\Lambda 
 (\Sigma -\sqrt{2}\partial_y)e^{-\Lambda}$ 
 under the gauge transformation. 
However, if we consider the non-local 
 operator, $\phi={\mathcal P}\exp( \int \Sigma dy)$,  
 as the {\rm new} Higgs doublet, 
 it can have the gauge invariant Yukawa 
 interactions 
 with the wall-localized 
 quarks and leptons\cite{Hall:2001zb}. 
Another possibility is 
 to consider the situation 
 where quarks and leptons are 
 the bulk fields\cite{Burdman:2002se},
 where the Yukawa interactions 
 is originated in the 5D gauge interactions.

%%%%%%%%%%%%%%%%%%%%%%%%%%%%%%%%%%%%%%%%%%%%%%%%%%%%%%
\vskip 1.5cm

\leftline{\bf Acknowledgments}

N.H. would like to thank C. S. Lim for a lot of very helpful 
 discussions. 
T.Y. would like to thank S. Teraguchi for useful discussions, 
 and is supported by a Grant-in-Aid for 
 the 21st Century COEhCenter for Diversity 
 and Universality in Physicsh. 
This work was supported in part by  Scientific Grants from 
 the Ministry of Education and Science, 
 Grant No.\ 14039207, 
 Grant No.\ 14046208, \ Grant No.\ 14740164 (N.H.),
 Grant No.\ 13135215,
 Grant No.\ 13640284 (Y.H.),
 Grant No.\ 13135217,
 Grant No.\ 15340078 (Y.K.). 

\vspace{5mm}
\leftline{\bf Note added}

\par
After this work was completed we noticed the work of 
 Scrucca, Serone and Silvestrini\cite{25}, where a
 similar idea is considered in non-SUSY $SU(3)_c \times SU(3)_W$. 
They estimated the effects to the effective potential 
 induced from bulk and wall-localized fields mixings and
 wall-localized kinetic terms.   
And they also consider flavour symmetry breaking 
 through Yukawa interactions with  
 the non-local operator induced by integrating out
 heavy bulk fields\cite{26}. 
The difference between their scenario and ours
 is that the electro-weak symmetry
 breaking is realized by the effect of  
 wall-localized fields 
 in the former, while 
 by the effect of 
 bulk fields possessing 
 degrees of freedom of $\eta$ and $\eta'$
 in the latter.

\vskip 1.5cm

\appendix
\section{The derivation of $V_{eff}^{gauge}$ in $SU(6)$}

The one loop effective potential 
 of the Wilson line degrees of freedom of  
 $A_5$ is given by 
\begin{eqnarray}
& & V_{{\rm{eff}}}[A^0] = 
            - (D-2){i \over 2}\mbox{Trln} D_{M}^{0} D^{0M} , 
\label{Veff}
\end{eqnarray}
where
$
D_L(A^0)D^L(A^0)
= \partial_\mu \partial^\mu + D_y(A^0)D^y(A^0),
$
with
$
A^0 \equiv \langle A_5 \rangle = 
   {1\over g R}\sum_a a_a {\lambda_a \over 2}.
$
We can always take the base of $\langle A_5 \rangle=0$
 by the gauge transformation, 
$
\Omega(x^\mu,y)=e^{i\sum a_a {\lambda_a\over 2} {y\over R}},
$
as 
\begin{eqnarray}
&&\langle A_5 \rangle  \rightarrow  
 \Omega(y)\langle A_5 \rangle\Omega(y)^\dagger
 -{i\over g}\Omega(y)\partial_y \Omega(y)^\dagger 
= \langle A_5 \rangle 
    -{1\over gR}\sum a_a {\lambda_a\over 2}=0 .
\end{eqnarray}
In this base the parities are given as 
\begin{eqnarray}
P &\rightarrow& P = \Omega(-y)P\Omega(y)^\dagger ,  \\
P' &\rightarrow& P' = \Omega(\pi R-y)P'\Omega(\pi R +y)^\dagger .
\end{eqnarray}
Then $P$ and $P$ correspond to the Wilson loop $W_C$.

Introducing a fluctuation field 
 $B \equiv \sum_a B_a {\lambda_a \over 2}$, 
 the effective potential 
 is given by 
\begin{eqnarray}
&&{\rm Tr ln}B D_y(A^0)D_y(A^0) B 
=-{\rm Tr}
(\sum_c [\partial_y B_c
-{1\over R}\sum_{a,b}f_{abc}a_aB_b]{\lambda_c\over 2})^2.
\label{35}
\end{eqnarray}
As is shown in the section 3, 
 we can always take VEV as 
$
a_{16} =a,
$
and other $a_i =0$, 
 by using the residual $SU(2)\times U(1)$ {\it global} symmetry. 
Thus, only structure constants relating 
 $\lambda_{16}$ are needed for the 
 calculation of Eq.(\ref{35}).
They are given by  
\begin{eqnarray}
&& f_{1,16,19}=f_{2,16,18}=f_{3,16,17}=f_{4,16,21}=f_{5,16,20}= \nonumber\\
&& f_{9,16,23}=f_{10,16,22}=f_{16,25,34}=f_{16,33,26}={1\over 2},\nonumber\\
&& f_{8,16,17}={1\over 2\sqrt{3}},\;\;\;\;\;\;
 f_{15,16,17}={1\over 2\sqrt{6}},\;\;\;\;\;\;
 f_{16,17,24}={\sqrt{10}\over 4}.
\end{eqnarray}
Where
 the generators are numbered as
\begin{eqnarray}
&&\left(
\begin{array}{cccccc}
     &(1,2)&(4,5)&(9,10)&(16,17)&(25,26)\\
(1,2)&   &(6,7)&(11,12)&(18,19)&(27,28)\\
(4,5)&(6,7)  &   &(13,14)&(20,21)&(29,30)\\
(9,10)&(11,12)&(13,14) &   &(22,23)&(31,32)\\
(16,17)&(18,19)&(20,21)&(22,23)&     &(33,34)\\
(25,26)&(27,28)&(29,30)&(31,32)&(33,34)& 
\end{array}
\right).
\end{eqnarray}
{}For examples, 
$(1,2)$ stands for 
\begin{eqnarray}
\lambda_1=\left(
\begin{array}{cccccc}
  & 1 & & & &  \\
 1 &  & & & &  \\
  &  & & & &  \\
  &  & & & &  \\
  &  & & & &  \\
  &  & & & &  
\end{array}
\right),\;\;\lambda_2=\left(
\begin{array}{cccccc}
  & -i & & & &  \\
 i &  & & & &  \\
  &  & & & &  \\
  &  & & & &  \\
  &  & & & &  \\
  &  & & & &  
\end{array}
\right).
\end{eqnarray}
The diagonal generators are 
\begin{eqnarray}
&&\lambda_3 = \diag(1,-1,0,0,0,0), \;\;\;\;\;
  \lambda_8 = {1\over \sqrt{3}}\diag(1,1,-2,0,0,0), \nonumber \\
&&\lambda_{15} = {1\over \sqrt{6}} \diag(1,1,1,-3,0,0), \;\;\;\;\;
  \lambda_{24} = {1\over \sqrt{10}}\diag(1,1,1,1,-4,0), \nonumber \\
&&\lambda_{35} = {1\over \sqrt{15}}\diag(1,1,1,1,1,-5), 
\end{eqnarray}
Then, Eq.(\ref{35}) becomes 
\begin{eqnarray}
&&{1\over 2}[(\partial_y B_i)^2 
+(\partial_y B_j -{1\over 2R}aB_k)^2+(\partial_y B_k +{1\over 2R}aB_j)^2
   \nonumber \\
&& +(\partial_y B_3 -{1\over 2R}aB_{17})^2
+(\partial_y B_8 -{1\over 2\sqrt{3}R}aB_{17})^2\nonumber\\
&&+(\partial_y B_{15} -{1\over 2\sqrt{6}R}aB_{17})^2 
  +(\partial_y B_{24} -{\sqrt{10}\over 4R}aB_{17})^2 \nonumber\\ 
&&+(\partial_y B_{17} +{1\over 2R}aB_3 +{1\over 2\sqrt{3}R}aB_8
         +{1\over 2\sqrt{6}R}aB_{15}+{\sqrt{10}\over 4R}aB_{24})^2],
\label{1}
\end{eqnarray}
where $i=6,7,11,12,13,14,16,27,28,29,30,31,32,35$, and
 $(j,k)=(1,19)$,$(2,18)$, $(4,21)$,
 $(5,20)$,
 $(9,23),(10,22),(25,34),(26,33)$. 
In $i=6,7,11,12,13,14,35$ in Eq.(\ref{1}), 
 eigenvalues are 
${n^2 \over R^2}$ and ${(n+1)^2 \over R^2}$ 
for 
 $(+,+)$ states of $A_\mu$ and 
 $(-,-)$ states of $A_5$, respectively. 
In $i=16$ in Eq.(\ref{1}), 
 eigenvalue is 
 ${(n+1)^2 \over R^2}$ and ${n^2 \over R^2}$ for 
 $(-,-)$ state of $A_\mu$ and 
 $(+,+)$ state of $A_5$, respectively.  
In $i=27,28,29,30,31,32$ in Eq.(\ref{1}), 
 eigenvalues are 
 ${(n+1/2)^2\over R^2}$ and ${(n+1/2)^2 \over R^2}$ for 
 $(-,+)$ states of $A_\mu$ and 
 $(+,-)$ states of $A_5$, respectively.  
In $(j,k)=(1,19)$, $(2,18)$, $(4,21)$, $(5,20)$, $(9,23)$, $(10,22)$, 
 eigenvalues are 
 ${(n+1/2+a/2)^2 \over R^2}$, ${(n+1/2-a/2)^2 \over R^2}$ and
 ${(n+1/2+a/2)^2 \over R^2}$, ${(n+1/2-a/2)^2 \over R^2}$ for 
 $((+,-),(-,+))$ states of $A_\mu$ and 
 $((-,+),(+,-))$ states of $A_5$, respectively.  
In $(j,k)=(25,34),(26,33)$, 
 eigenvalues are 
 ${(n+a/2)^2 \over R^2}$, ${(n-a/2)^2 \over R^2}$ and
 ${(n+a/2)^2 \over R^2}$, ${(n-a/2)^2 \over R^2}$ for 
 $((-,-),(+,+))$ states of $A_\mu$ and 
 $((+,+),(-,-))$ states of $A_5$, respectively.  
As for $(B_3,B_8,B_{15},B_{24},B_{17})$,
the relevant part in Eq. (59) is simplified as
\begin{eqnarray}
&&{1\over 2}\bigg\{
\Big(\partial_y C_1 - {a \over R} B_{17} \Big)^2
+ \Big(\partial_y B_{17} + {a \over R} C_1 \Big)^2 \nonumber \\
&& ~~~~~~~~~ +(\partial_y C_2 )^2
+(\partial_y C_3 )^2 + (\partial_y C_4 )^2 \bigg\} ~~,
\label{Cs}
\end{eqnarray}
in an appropriate basis.
Here new fields $C_j$s are introduced by
\begin{eqnarray}
C_1 \Lambda_1 + C_2 \Lambda_2 + C_3 \Lambda_3 + C_4 \Lambda_4
= B_3 \lambda_3 + B_8 \lambda_8 + B_{15} \lambda_{15} + B_{24} \lambda_{24}
\label{Cfield1}
\end{eqnarray}
by use of new bases $\Lambda_j$s with diagonal elements.
For instance, $C_1$ and $\Lambda_1$ are given by
\begin{eqnarray}
C_1 = {1\over 2}B_3 +{1\over 2\sqrt{3}}B_8
         +{1\over 2\sqrt{6}}B_{15}+{\sqrt{10}\over 4}B_{24}
\label{C1}
\end{eqnarray}
and
\begin{eqnarray}
\Lambda_1 = \mbox{diag}(1, 0, 0, 0, -1, 0) ~~,
\end{eqnarray}
respectively.
{}From the above Eq. (\ref{Cs}), the eigenvalues are given by
  ${(n+a)^2 \over R^2}$, ${(n-a)^2 \over R^2}$,
${n^2 \over R^2}$, ${n^2 \over R^2}$ and ${n^2 \over R^2}$.

Then, 
 the effective potential for gauge and ghost 
 is given by 
\begin{eqnarray}
V_{eff}^{g+gh}
&&= -2 {i\over 2}\int {d^4p\over (2\pi)^4}{1\over 2\pi R}
[\sum_i
 \sum_{n=0}^{\infty}{\rm ln}(-p^2+{n^2\over R^2}) 
+\sum_j 
 \sum_{n=1}^{\infty}{\rm ln}(-p^2+{n^2\over R^2}) \nonumber \\
&& +\sum_k 
    \sum_{n=0}^{\infty}{\rm ln}(-p^2+{(n+1/2)^2\over R^2}) 
     \nonumber\\
&& +\sum_{(l,m)} 
    \sum_{n=0}^{\infty}{\rm ln}[
     (-p^2+{(n+1/2+a/2)^2\over R^2})+ (-p^2+{(n+1/2-a/2)^2\over R^2})] 
 \nonumber \\
&& +\sum_{(p,q)} \sum_{n=0}^{\infty}{\rm ln}(-p^2+{(n+a/2)^2\over R^2})+ 
    \sum_p \sum_{n=1}^{\infty}{\rm ln}(-p^2+{(n-a/2)^2\over R^2}) \\
&& +3\sum_{n=0}^{\infty}{\rm ln}(-p^2+{n^2\over R^2}) 
 +\sum_{n=0}^{\infty}{\rm ln}
     (-p^2+{(n+a)^2\over R^2})+ 
   \sum_{n=1}^{\infty}{\rm ln}
     (-p^2+{(n-a)^2\over R^2}) ],  \nonumber 
\end{eqnarray}
and 
 for $A_5$ part 
 is given by 
\begin{eqnarray}
V_{eff}^{A_5}
&&= -1 {i\over 2}\int {d^4p\over (2\pi)^4}{1\over 2\pi R}
[\sum_i
 \sum_{n=0}^{\infty}{\rm ln}(-p^2+{n^2\over R^2}) 
+\sum_j 
 \sum_{n=1}^{\infty}{\rm ln}(-p^2+{n^2\over R^2}) \nonumber \\
&& +\sum_k 
    \sum_{n=0}^{\infty}{\rm ln}(-p^2+{(n+1/2)^2\over R^2}) 
     \nonumber\\
&& +\sum_{(l,m)} 
    \sum_{n=0}^{\infty}{\rm ln}[
     (-p^2+{(n+1/2+a/2)^2\over R^2})+ (-p^2+{(n+1/2-a/2)^2\over R^2})] 
 \nonumber \\
&& +\sum_{(p,q)} \sum_{n=0}^{\infty}{\rm ln}(-p^2+{(n+a/2)^2\over R^2})+ 
    \sum_p \sum_{n=1}^{\infty}{\rm ln}(-p^2+{(n-a/2)^2\over R^2}) \\
&& +3\sum_{n=0}^{\infty}{\rm ln}(-p^2+{n^2\over R^2}) 
 +\sum_{n=0}^{\infty}{\rm ln}
     (-p^2+{(n+a)^2\over R^2})+ 
   \sum_{n=1}^{\infty}{\rm ln}
     (-p^2+{(n-a)^2\over R^2}) ],  \nonumber 
\end{eqnarray}
where $i=6,7,11,12,13,14,35$, 
 $j=16$, 
 $k=27,28,29,30,31,32$, 
 $(l,m)=(1,19)$, $(2,18)$, $(4,21)$, $(5,20)$, $(9,23)$, $(10,22)$, 
 $(p,q)=(25,34),(26,33)$, and 
 the last line terms come from 
 $(B_3,B_8,B_{15},B_{24},B_{17})$.  
We omit terms without $a$ dependences, 
 since they have nothing to
 do with the dynamics of determining $a$. 
Then, we can obtain the VEV dependent effective potential as
\begin{eqnarray}
V_{eff}^{g+gh+A_5}
&=& -3 {i\over 2}\int {d^4p\over (2\pi)^4}{1\over 2\pi R}[ 
 6 \sum_{n=-\infty}^{\infty}{\rm ln}
           (-p^2+{(n+1/2-a/2)^2\over R^2}) \nonumber \\ 
&&+ 2 \sum_{n=-\infty}^{\infty}{\rm ln}(-p^2+{(n-a/2)^2\over R^2}) +
     \sum_{n=-\infty}^{\infty}{\rm ln}(-p^2+{(n-a)^2\over R^2})] \\
&=& -{3\over2}C\sum_{n=1}^{\infty}{1\over n^5}
    [6 \cos (\pi n ({a}-{1}))+2\cos(\pi na)+\cos(2\pi na)],
\end{eqnarray}
where 
$C ={3\over 64\pi^7R^5}$.

The SUSY version of the effective potential 
 is easily obtained from 
 this non-SUSY one. 
As in Ref.\cite{HHHK}, 
 it is obtained by replacing 
 the coefficient ${-2/3}C \rightarrow -2C$
 due to the change of degrees of freedom, 
 and adding the factor $(1-\cos(2\pi n\beta))$ 
 to $V_{eff}^{gauge}$,
 where $\beta$ is the 
 parameter of SS SUSY breaking.

\section{The derivation of $V_{eff}^{m}$}

The effective potential induced from 
 bulk fields 
 is given by  
\begin{eqnarray}
\label{60}
&~& V_{\mbox{eff}}[A^0]^{fermion} = 
               f(D){i \over 2}\mbox{Trln} D_{L}^{0} D^{0L} ,\\
&~& V_{\mbox{eff}}[A^0]^{scalar} = 
               - 2 {i \over 2}\mbox{Trln} D_{L}^{0} D^{0L} ,
\label{61}
\end{eqnarray}
where $f(D) = 2^{[D/2]}$.  
The $V_{eff}^m$ can be calculated similar to
 the $V_{eff}^{gauge}$. 
We introduce a fluctuation field $B$ according 
 to the representation of bulk fields in 
 Eqs.(\ref{60}) and (\ref{61}) as Eq.(\ref{35}), and 
 calculating the eigenvalues. 
Here, we show only eigenvalues 
 both in $SU(3)_c\times SU(3)_W$ and $SU(6)$ models.

In $SU(3)_c\times SU(3)_W$ model, 
 adjoint representation fields with $\eta \eta'=+$ have 
 eigenvalues, $2\times{n^2 \over R^2}$, 
 ${(n \pm a)^2 \over R^2}$, and 
 $2\times {(n \pm a/2)^2 \over R^2}$. 
Thus, $N_a^{(+)}$ numbers of Dirac fermion 
 induce $2N_a^{(+)}C \sum {1\over n^5}$ 
 $[\cos(2\pi na)+2\cos(\pi na)]$.  
The adjoint representation fields 
 with $\eta \eta'=-$ have 
 eigenvalues, $2\times{(n+1/2)^2 \over R^2}$, 
 ${(n \pm a+1/2)^2 \over R^2}$, and 
 $2\times {(n \pm a/2+1/2)^2 \over R^2}$. 
Thus, $N_a^{(-)}$ numbers of Dirac fermion 
 induce $2N_a^{(-)}C \sum {1\over n^5}
 [\cos(2\pi n(a-1/2))+2\cos(\pi n(a-1))]$.  
The fundamental 
 representation fields with $\eta \eta'=+$ have 
 eigenvalues, ${n^2 \over R^2}$ and 
 ${(n \pm a)^2 \over R^2}$. 
Then $N_f^{(+)}$ and $N_s^{(+)}$ numbers of Dirac fermion 
 and complex scalar 
 induce $(2N_f^{(+)}-N_s^{(+)})C \sum 
 {1\over n^5}\cos(\pi na)$.  
The fundamental 
 representation fields with $\eta \eta'=-$ have 
 eigenvalues, ${(n+1/2)^2 \over R^2}$ and 
 ${(n \pm a +1/2)^2 \over R^2}$. 
Then $N_f^{(-)}$ and $N_s^{(-)}$ numbers of Dirac fermion 
 and complex scalar 
 induce $(2N_f^{(-)}-N_s^{(-)})C \sum 
 {1\over n^5}\cos(\pi n(a-1))]$.

In $SU(6)$ model, 
 adjoint representation fields with $\eta \eta'=+$ have 
 eigenvalues, $11\times{n^2 \over R^2}$, 
 $6\times{(n+1/2)^2 \over R^2}$, 
 ${(n \pm a)^2 \over R^2}$, 
 $2\times {(n \pm a/2)^2 \over R^2}$, and 
 $6\times {(n \pm a/2+1/2)^2 \over R^2}$.  
Thus, $N_a^{(+)}$ numbers of Dirac fermion 
 induce $2N_a^{(+)}C \sum {1\over n^5}$ 
 $[6\cos(\pi n(a-1))+2\cos(\pi na)+\cos(2\pi na)]$.  
The adjoint representation fields 
 with $\eta \eta'=-$ have 
 eigenvalues, $11\times{(n+1/2)^2 \over R^2}$, 
 $6\times{n^2 \over R^2}$, 
 ${(n \pm a+1/2)^2 \over R^2}$, 
 $2\times {(n \pm a/2+1/2)^2 \over R^2}$, and 
 $6\times {(n \pm a/2)^2 \over R^2}$.  
Thus, $N_a^{(-)}$ numbers of Dirac fermion 
 induce $2N_a^{(-)}C \sum {1\over n^5}$ 
 $[6\cos(\pi na)+2\cos(\pi n(a-1))+\cos(2\pi n(a-1/2))]$.  
The fundamental 
 representation fields with $\eta \eta'=+$ have 
 eigenvalues, ${n^2 \over R^2}$, 
 $3\times{(n+1/2)^2 \over R^2}$, and 
 ${(n \pm a)^2 \over R^2}$. 
Then $N_f^{(+)}$ and $N_s^{(+)}$ numbers of Dirac fermion 
 and complex scalar 
 induce $(2N_f^{(+)}-N_s^{(+)})C \sum 
 {1\over n^5}\cos(\pi na)$.  
The fundamental 
 representation fields with $\eta \eta'=-$ have 
 eigenvalues, ${(n+1/2)^2 \over R^2}$, 
 $3\times{n^2 \over R^2}$, and 
 ${(n \pm (a-1))^2 \over R^2}$. 
Then $N_f^{(-)}$ and $N_s^{(-)}$ numbers of Dirac fermion 
 and complex scalar 
 induce $(2N_f^{(-)}-N_s^{(-)})C \sum 
 {1\over n^5}\cos(\pi n(a-1))]$.

The effective potential 
 in the SUSY case with SS breaking 
 can also be obtained as in Ref.\cite{HHHK}. 
We add the factor $(1-\cos(2\pi n\beta ))$ 
 in the effective potential induced from the 
 fermion (ordinary particle) 
 contributions.

\vspace{1cm}

% A useful Journal macro
\def\jnl#1#2#3#4{{#1}{\bf #2} (#4) #3}

\def\Zphys{{\em Z.\ Phys.} }
\def\jssc{{\em J.\ Solid State Chem.\ }}
\def\jpsJ{{\em J.\ Phys.\ Soc.\ Japan }}
\def\ptps{{\em Prog.\ Theoret.\ Phys.\ Suppl.\ }}
\def\PTP{{\em Prog.\ Theoret.\ Phys.\  }}

\def\JMP{{\em J. Math.\ Phys.} }
\def\NPB{{\em Nucl.\ Phys.} B}
\def\NP{{\em Nucl.\ Phys.} }
\def\PLB{{\em Phys.\ Lett.} B}
\def\PL{{\em Phys.\ Lett.} }
\def\PRL{\em Phys.\ Rev.\ Lett. }
\def\PRB{{\em Phys.\ Rev.} B}
\def\PRD{{\em Phys.\ Rev.} D}
\def\PRe{{\em Phys.\ Rep.} }
\def\AP{{\em Ann.\ Phys.\ (N.Y.)} }
\def\RMP{{\
em Rev.\ Mod.\ Phys.} }
\def\ZPC{{\em Z.\ Phys.} C}
\def\SCI{\em Science}
\def\CMP{\em Comm.\ Math.\ Phys. }
\def\MPLA{{\em Mod.\ Phys.\ Lett.} A}
\def\IJMPA{{\em Int.\ J.\ Mod.\ Phys.} A}
\def\IJMPB{{\em Int.\ J.\ Mod.\ Phys.} B}
\def\EPJC{{\em Eur.\ Phys.\ J.} C}
\def\PR{{\em Phys.\ Rev.} }
\def\JHEP{{\em JHEP} }
\def\cmp{{\em Com.\ Math.\ Phys.}}
\def\JPA{{\em J.\  Phys.} A}
\def\CQG{\em Class.\ Quant.\ Grav. }
\def\ATMP{{\em Adv.\ Theoret.\ Math.\ Phys.} }
\def\ibid{{\em ibid.} }

\leftline{\bf References}

\renewenvironment{thebibliography}[1]
         {\begin{list}{[$\,$\arabic{enumi}$\,$]}  % {\arabic{enumi}.}
         {\usecounter{enumi}\setlength{\parsep}{0pt}
          \setlength{\itemsep}{0pt}  \renewcommand{\baselinestretch}{1.2}
          \settowidth
         {\labelwidth}{#1 ~ ~}\sloppy}}{\end{list}}

\end{document}